\documentclass[twocolumn,prl,showpacs,floatfix]{revtex4}
\usepackage{graphicx}
\usepackage{epsfig}
\usepackage{rotating}
\usepackage{amsmath}
\usepackage{amsfonts}
\usepackage{amssymb}
\usepackage{enumerate}
\usepackage{longtable}
\usepackage{dcolumn}
\usepackage{bm}

\begin{document}

\title{Ground state properties, excitation spectra and phase transitions in 
the $S=1/2$ and $S=3/2$ bilayer Heisenberg models on the honeycomb Lattice}

\author{J. Oitmaa }
\affiliation{School of Physics, The University of New South Wales,
Sydney 2052, Australia}

\author{R. R. P. Singh}
\affiliation{University of California Davis, CA 95616, USA}

\date{\rm\today}

\begin{abstract}
Motivated by the observation of a disordered spin ground state in the
$S=3/2$ material Bi$_3$Mn$_4$O$_{12}$NO$_3$, we study the ground state
properties and excitation spectra of the $S=3/2$ 
(and for comparison $S=1/2$) bilayer
Heisenberg model on the honeycomb lattice, with and without frustrating
further neighbor interactions. We use series expansions 
around the N\'eel state to calculate properties of the magnetically
order phase. Furthermore, series expansions in $1/\lambda=J_1/J_{\perp}$,
where $J_1$ is an in-plane exchange constant and $J_\perp$ is the
exchange constant between the layers are used to study properties
of the spin singlet phase. For the unfrustrated case,
our results for the phase transitions
are in very good agreement with recent Quantum Monte Carlo studies.
We also obtain the excitation spectra in the disordered phase
and study the change in the critical $\lambda$ when frustrating
exchange interactions are added to the $S=3/2$ system and find a rapid
suppression of the ordered phase with frustration. Implications
for the material Bi$_3$Mn$_4$O$_{12}$NO$_3$ are discussed.

\end{abstract}

\pacs{74.70.-b,75.10.Jm,75.40.Gb,75.30.Ds}

\maketitle

\section{Introduction}

The honeycomb lattice is a bipartite but low-coordination number
two-dimensional lattice.
Thus colinear antiferromagnetism is unfrustrated on the lattice.
However, it is more susceptible to disorder due to its low coordination
number. Recent interest in antiferromagnetism on the honeycomb lattice comes 
from many directions. First, theoretical studies of quantum spin models
on the lattice find a rich phase diagram, with several colinear,
spiral and spin-disordered phases.\cite{oitmaa78,rastelli,reger,weihong91,oitmaa92,mattsson,
fouet,mulder,cabra11,mosadeq,farnell,alb,oitmaa-singh,tsirlin} 
Second, a recent Quantum Monte Carlo finding
that the Hubbard model on the honeycomb lattice shows strong 
evidence for a quantum spin-liquid phase sandwiched between the
semi-metal phase at small $U/t$ and an ordered antiferromagnetic
phase at large $U/t$, has led to much
follow-up activity.\cite{meng,kpschmidt,paiva,clark,reuther} 
Third, the discovery of graphene and the search
for correlated topological insulator phases have also led to
interest in strongly correlated electron models on the honeycomb
lattice, including spin-orbit coupling.\cite{graphene,tinsulator} 

In this paper, our primary motivation comes from the bismuth manganese
oxynitrate material
Bi$_3$Mn$_4$O$_{12}$NO$_3$, an $S=3/2$ antiferromagnet, with 
a Curie-Weiss constant of order $250K$, which does not show any 
long range order down to $0.4 K$.\cite{smirnova} The material
consists of honeycomb lattices of $S=3/2$ $Mn$ spins which are
separated by bismuth and nitrate layers. Two such layers are separated
by bismuth atoms, forming a bilayer, and these bilayers are then
separated by significantly larger separations. Thus an appropriate
model for such a system is a spin-$3/2$ Heisenberg model on
the bilayer honeycomb lattice, with an in-plane nearest-neighbor exchange $J_1$
and an exchange between the bilayers of $J_\perp$.\cite{kandpal}
The bilayer exchange $J_\perp$ has been estimated by an
electronic structure calculation to be between
$J_1$ and $2 J_1$, whereas further neighbor in-plane exchanges,
which frustrate the system are down compared to $J_1$ by
an order of magnitude (less than $0.2 J_1$).\cite{kandpal}
This system has also been studied recently by Quantum Monte Carlo (QMC)
simulations and bond-operator based mean-field 
and variational theory\cite{ganesh,ganesh2}, which our results are compared to below.

We use series expansion methods to study the properties of 
the bilayer honeycomb Heisenberg model.\cite{book,advphy} Ising series expansions are
used to calculate the ground state energy and antiferromagnetic
order parameters in the N\'eel phase. Dimer series expansions
are used to calculate the ground state energy, triplet
energy gap and the excitation spectra in the disordered
singlet phase. The phase transition in this model
is known to be in the universality class of the $3d$
classical Heisenberg model.\cite{ganesh} We use this in our series
analysis to make our analysis more accurate. For the
nearest neighbor model, we find the phase transition to be located at
$J_\perp/J_1$ value of $1.66 \pm 0.01$ for $S=1/2$
an $9.34 \pm .20$ for $S=3/2$. These results are in very good
agreement with the values reported in the QMC study of
$1.645(1)$ and $9.194(3)$ for the two cases respectively.
The bond-operator theory\cite{ganesh2} is significantly less accurate. 

Since the material Bi$_3$Mn$_4$O$_{12}$NO$_3$ is unlikely to have
$J_\perp/J_1\approx 9$, frustration in the
plane must play a role in disordering the system.
Thus, we also study the model with frustration and discuss its
possible relevance to the materials. We also present results
for the excitation spectra in the disordered phase, which could
be helpful in further quantifying the experimental system.

\section{Series expansions for the honeycomb bilayer models}

We consider the Heisenberg model on the bilayer honeycomb lattice
with Hamiltonian
\begin{equation}
{\cal H}=J_1\sum_{<i,j>,a} \vec S_i^a\cdot \vec S_j^a
+J_2\sum_{<i,k>,a} \vec S_i^a\cdot \vec S_k^a
+J_\perp\sum_{i} \vec S_i^1\cdot \vec S_i^2.
\end{equation}
Hear $a=1,2$ denotes the spins in the two layers. The 
first sum runs over nearest-neighbors in the honeycomb planes
and the second sum is over the second neighbors in the honeycomb planes.
The third sum is over the neighboring spins between the two
bilayers.

For the unfrustrated models ($J_2=0$), we have calculated
Ising expansions for the ground state energy and sublattice
magnetization to $14$-th order for $S=1/2$ and to $12$-th order
for $S=3/2$. The dimer expansions in $1/\lambda=J_1/J_\perp$
are calculated for the ground state energy to order $12$ for $S=1/2$
and to order $9$ for $S=3/2$. The energy gap series is
calculated to order $10$ for $S=1/2$ and to order $6$ for $S=3/2$.
For $J_2/J_1$ non-zero, the energy gap series is calculated
in powers of $1/\lambda$ to order $6$ for $S=3/2$. The series
can be made available upon request.

\begin{figure}
\begin{center}
 \includegraphics[width=6cm,angle=270]{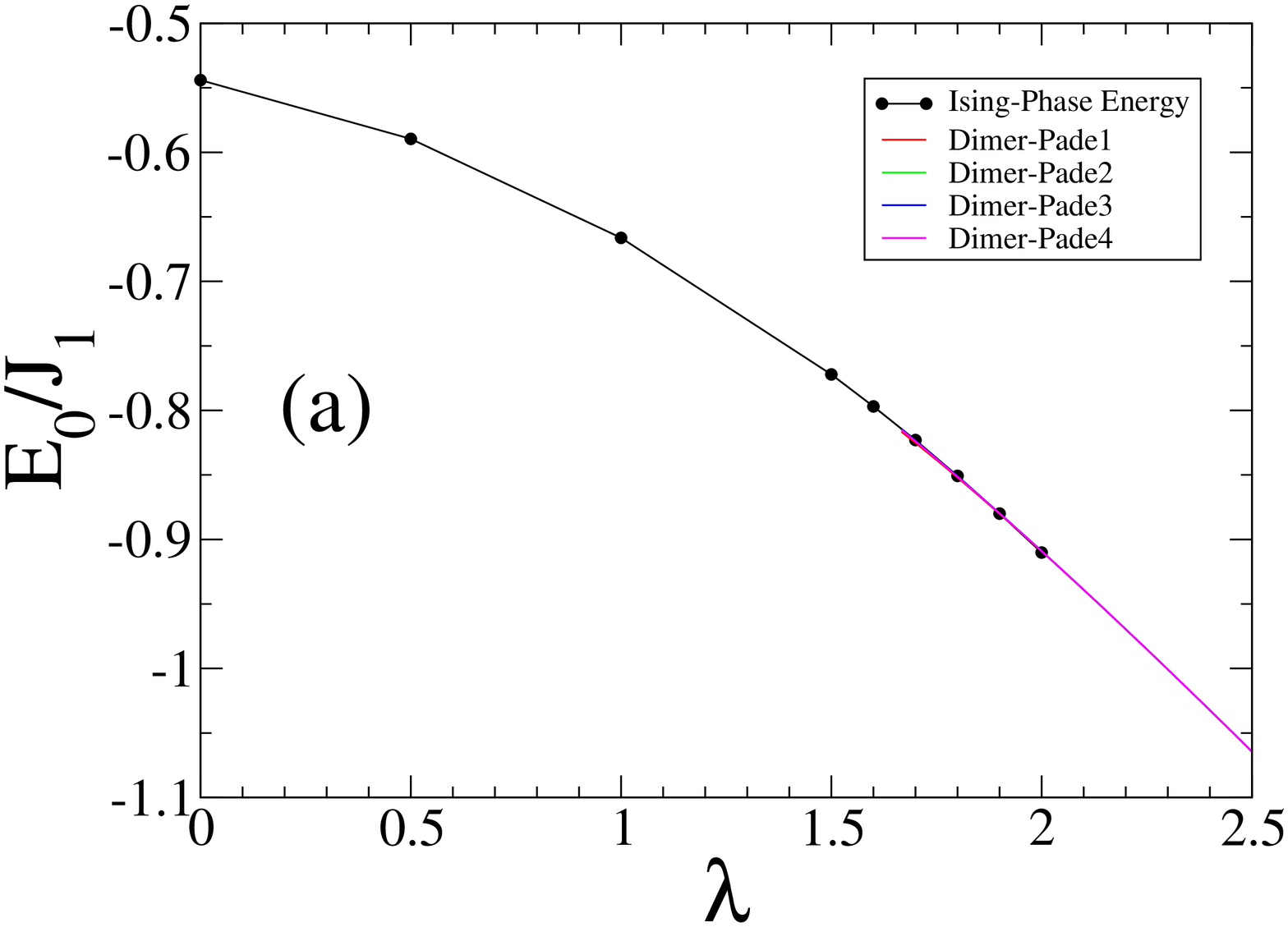}
 \includegraphics[width=6cm,angle=270]{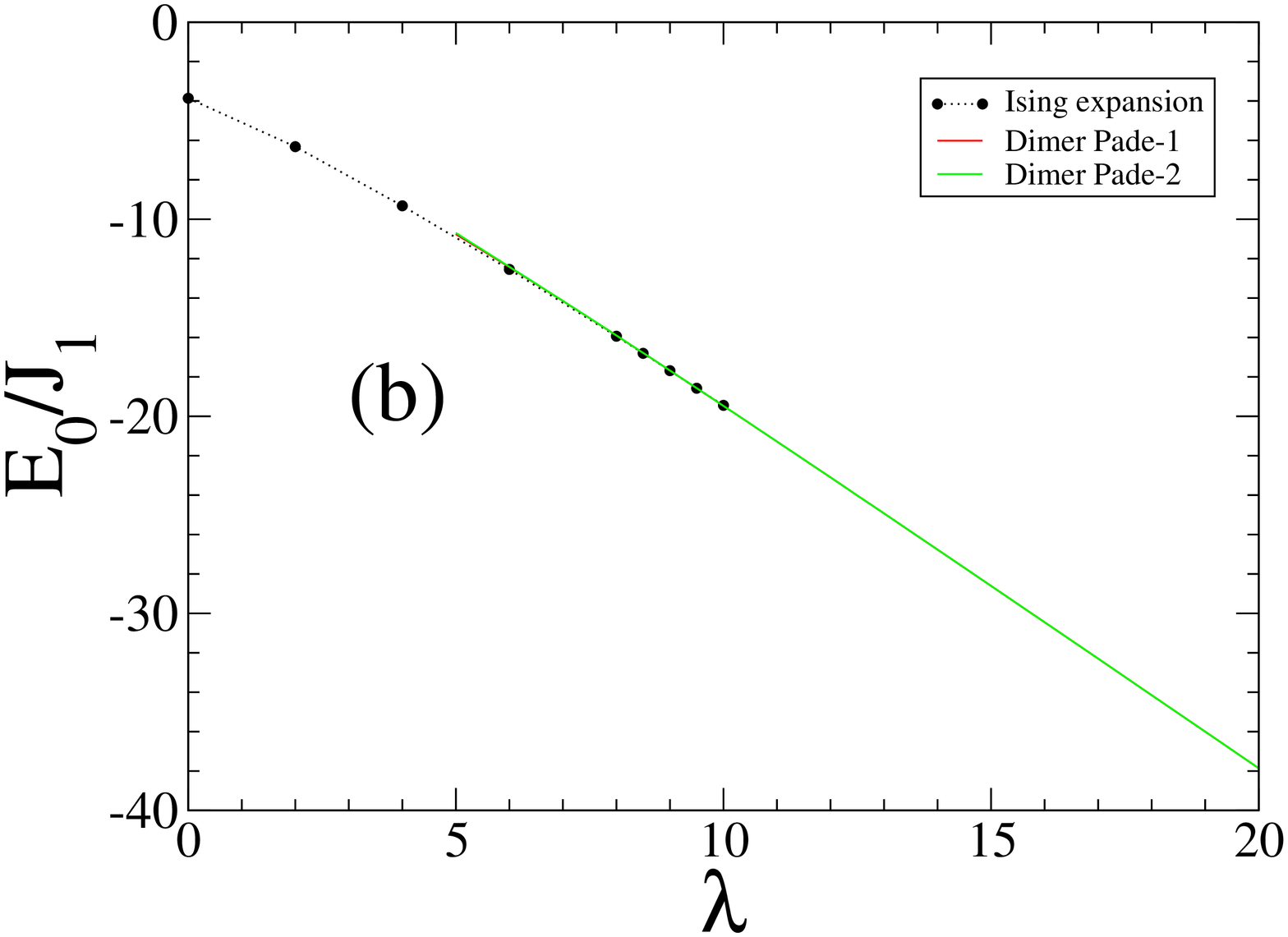}
\caption{\label{fig:Fig1} Ground state energy as a function of $\lambda=J_\perp/J_1$
for (a) $S=1/2$ and (b) $S=3/2$ models.
}
\end{center}
\end{figure}

In Fig.~1 (a) and (b), we show plots of the ground state energy for $J_2=0$.
For both $S=1/2$ and $S=3/2$,
the results from Ising and dimer expansions join smoothly as
expected for a second order phase transition. However, the energy is
not the best quantity to determine the location of the
phase transition. 

\begin{figure}
\begin{center}
 \includegraphics[width=6cm,angle=270]{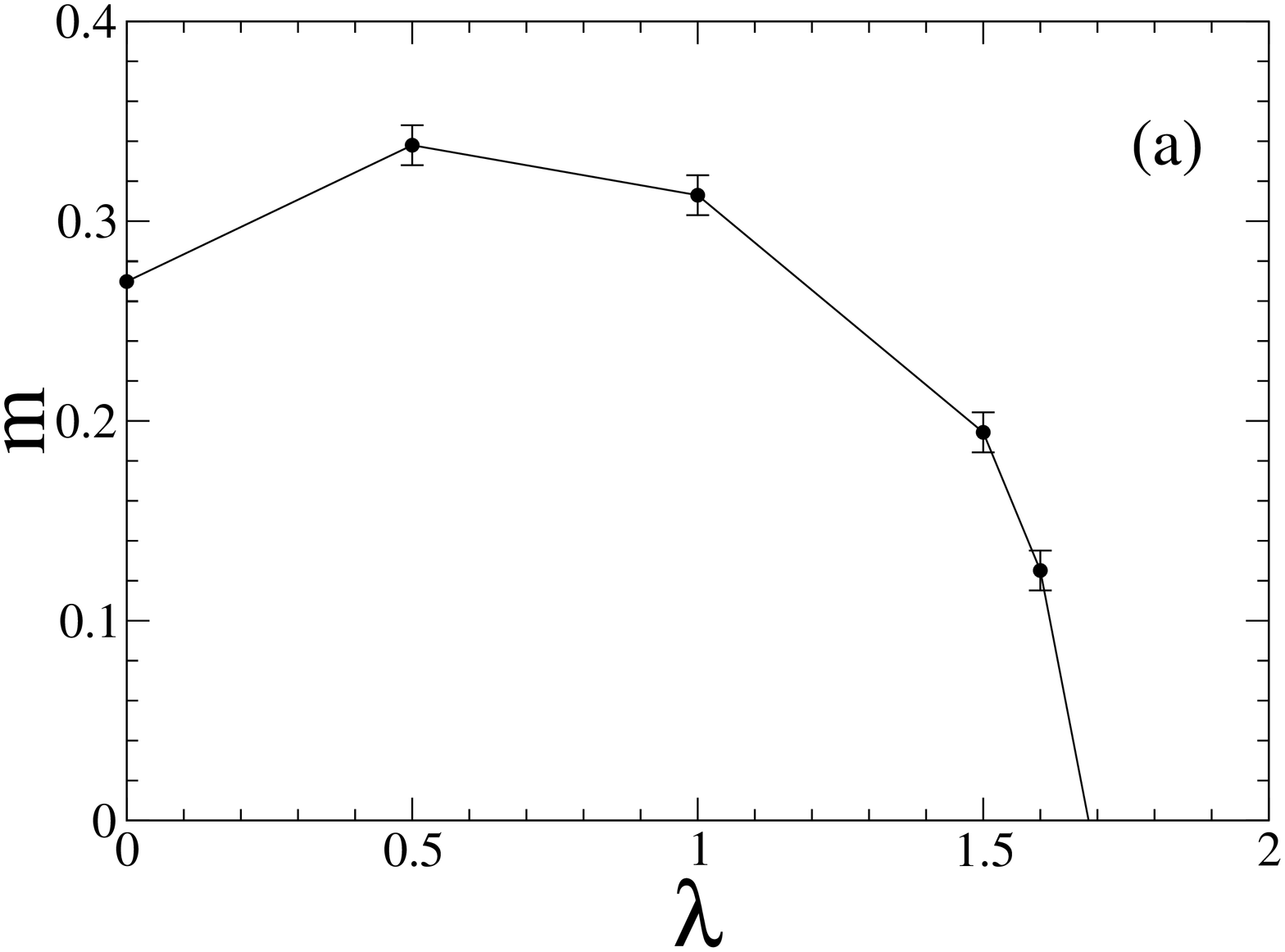}
 \includegraphics[width=6cm,angle=270]{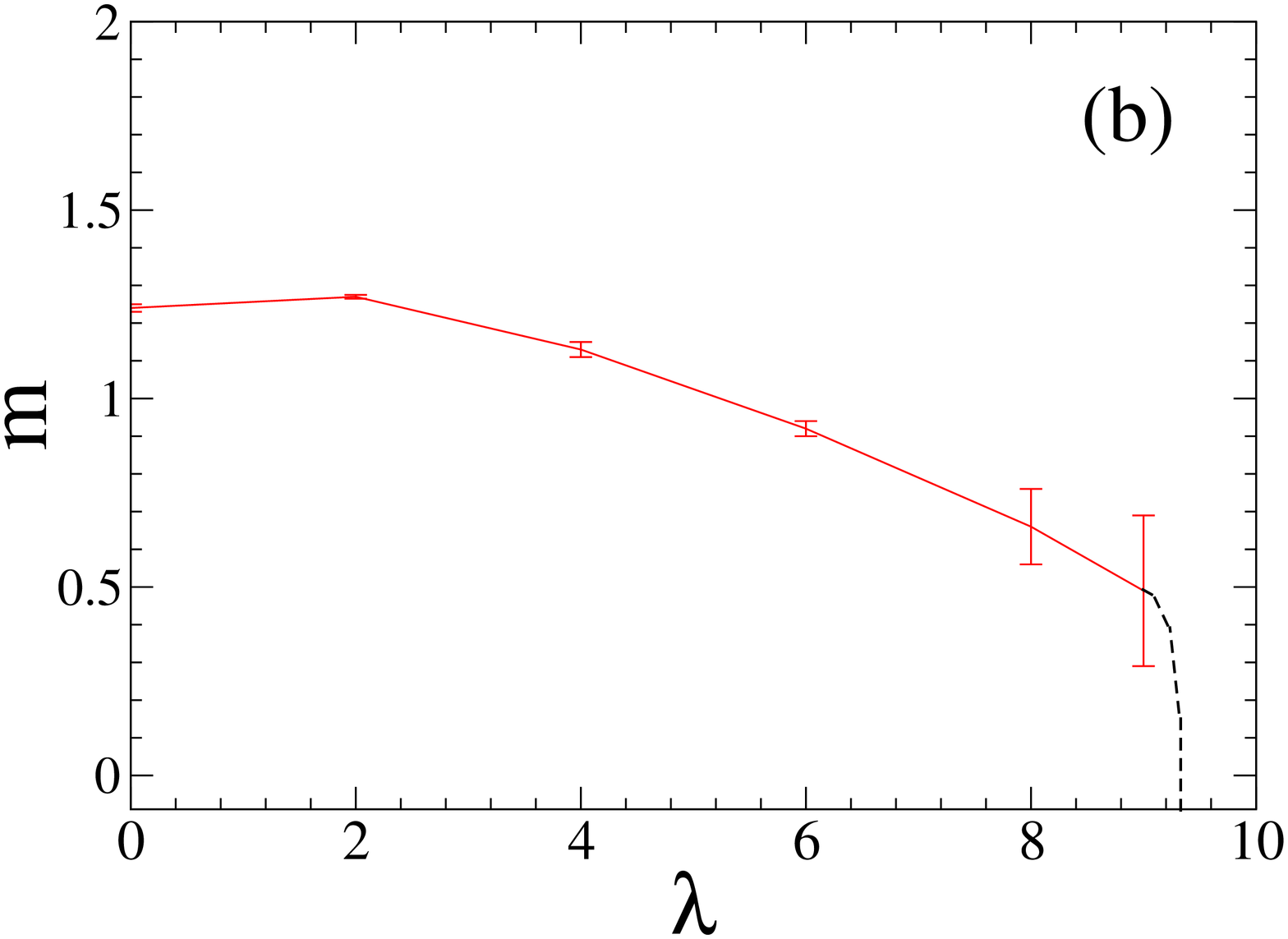}
\caption{\label{fig:Fig2} N\'eel order parameter as a function of $\lambda=J_\perp/J_1$
for (a) $S=1/2$ and (b) $S=3/2$ models.
}
\end{center}
\end{figure}

In Fig.~2 (a) and (b), we show the N\'eel order parameters for the $S=1/2$ and $S=3/2$ cases.
We use a square-root transformation\cite{huse} to remove the singularity
before analyzing the series by Pad\'e approximants.
Our analysis is not accurate close to the transition where the magnetization vanishes.
The uncertainties clearly become large for the shorter
$S=3/2$ series for $\lambda>8$ as the transition is approached. Beyond $\lambda=9$,
the rapid decrease is sketched in the plot by a dashed line.
The order parameter is also not the best way to get a precise estimate of
the transition point, because the series are not directly in the variable $1/\lambda$,
in which a power-law singularity, with an exponent $\beta$ is expected.
Rather, for every $\lambda$ the series is analyzed in the anisotropy variable to
calculate the order parameter for the Heisenberg model. It is difficult to
enforce the correct power-law singularity in such an analysis.

\begin{figure}
\begin{center}
 \includegraphics[width=7cm,angle=270]{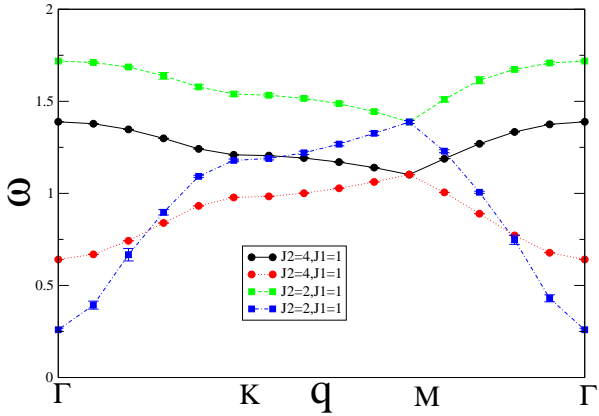}
\caption{\label{fig:Fig3} Triplon spectra for the
for $S=1/2$ models along the contour
$\Gamma(0,0), K({2\pi\over 3 a},0),
M({2\pi\over3a},{2\pi\over\sqrt{3}a}), \Gamma(0,0)$
in the Brillouin zone.
}
\end{center}
\end{figure}

\begin{figure}
\begin{center}
 \includegraphics[width=7cm,angle=270]{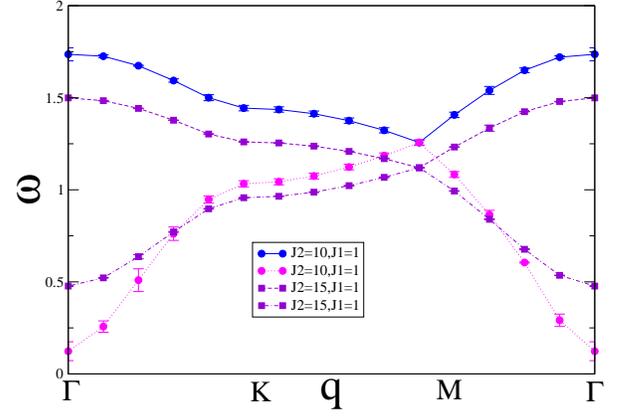}
\caption{\label{fig:Fig4} Triplon spectra for the
for $S=3/2$ models along the contour
$\Gamma(0,0), K({2\pi\over 3 a},0),
M({2\pi\over3a},{2\pi\over\sqrt{3}a}), \Gamma(0,0)$
in the Brillouin zone.
}
\end{center}
\end{figure}

The triplet excitation spectra for $S=1/2$ and $S=3/2$ along symmetry lines of
the Brillouin zone ($\Gamma(0,0), K({2\pi\over 3 a},0), 
M({2\pi\over3a},{2\pi\over\sqrt{3}a}), \Gamma(0,0)$)
are shown in Fig.~3 and Fig.~4 respectively. In the dimerized phase, there
are two branches of the triplet spectrum at every $k$ corresponding
to the two atoms per unit cell in the honeycomb lattice.
They become degenerate at the M point.
The spectral gap approaches zero as the transition to the N\'eel
phase is approached.

\begin{figure}
\begin{center}
 \includegraphics[width=6cm,angle=270]{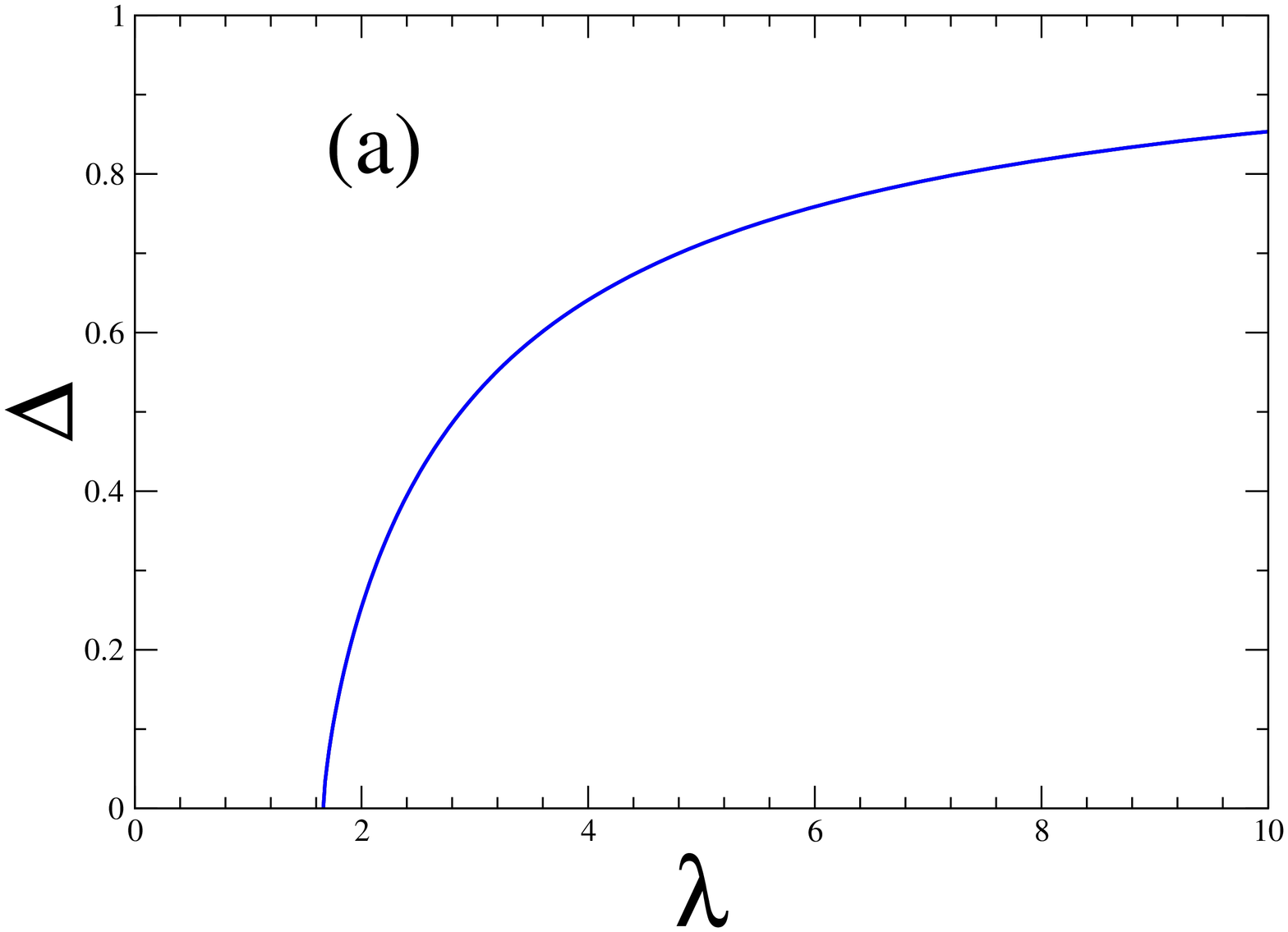}
 \includegraphics[width=6cm,angle=270]{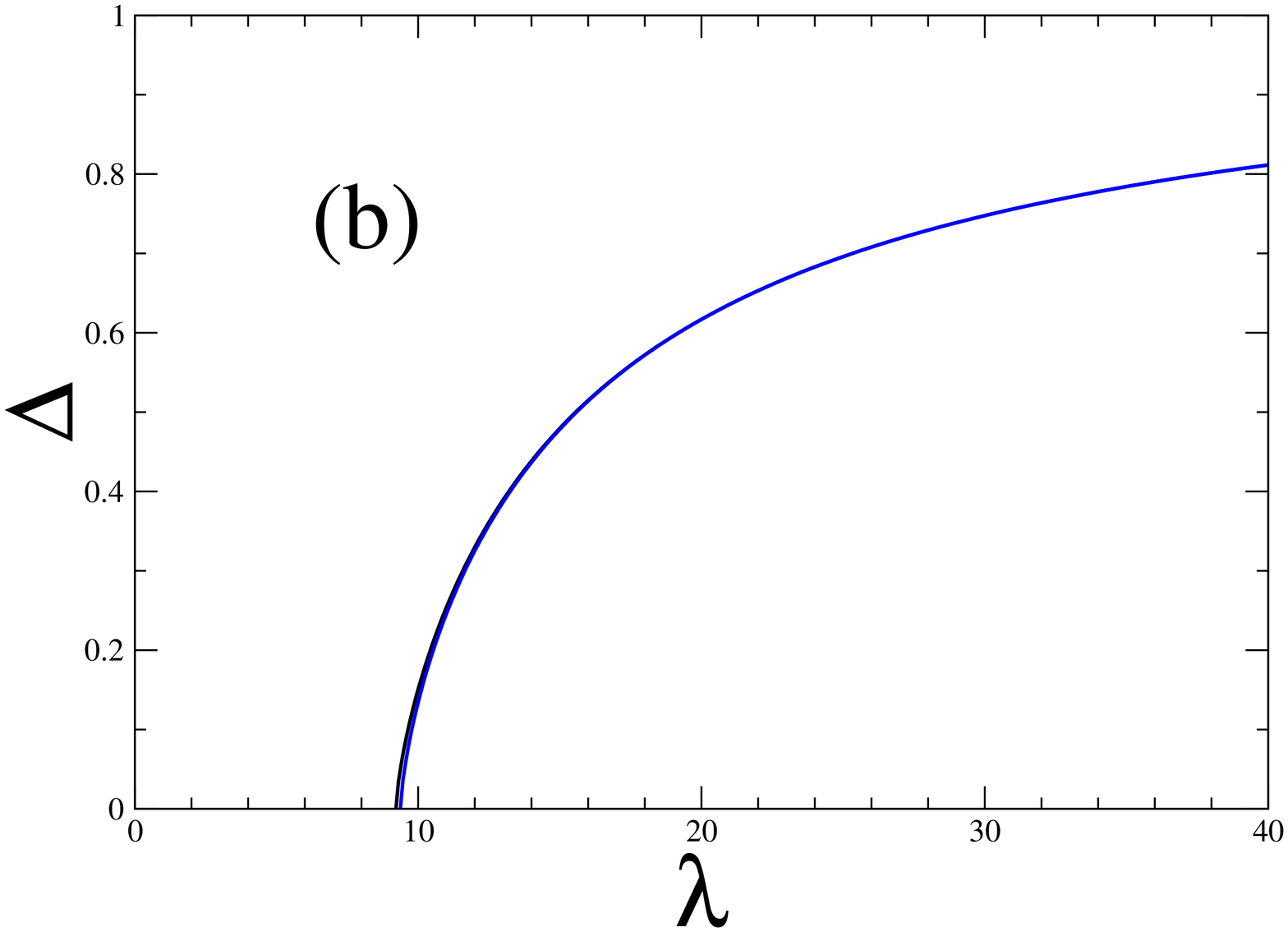}
\caption{\label{fig:Fig5} Energy gap as a function of $\lambda=J_\perp/J_1$
for (a) $S=1/2$ and (b) $S=3/2$ models.
}
\end{center}
\end{figure}

The most accurate way to determine the critical $\lambda$ 
in our study is
by analyzing the energy gap series. Here, we use the knowledge
that this series has a singularity at the critical $\lambda$
with a critical exponent $\nu$ given by the $3d$ Heisenberg
universal value $\nu=0.71$\cite{heisenberg}. Thus, we first raise the series to a
power $1/\nu \approx 1.408$, and then study it by Pad\'e 
approximants to see where it goes to zero. The Pad\'e approximants
show excellent internal consistency for the longer $S=1/2$
series. Three different approximants are shown in Fig.~5a and they
are almost indistinguishable from each other.
Thus the uncertainty comes primarily from varying
the value of $\nu$. We estimate $1/\lambda_c=0.6014\pm 0.002$,
which translates to $J_\perp/J_1=1.66\pm 0.02$. For $S=3/2$,
the shorter series though less accurate than for $s=1/2$ is still very accurate. Two different approximants
are shown in Fig.~5b, and one can see that they begin to
deviate a little close to the transition. In this case, we
estimate $1/\lambda_c=0.107\pm 0.002$, which translates to
$J_\perp/J_1=9.3\pm 0.2$. Both these results are in very good
agreement with the QMC study, which gives $J_\perp/J_1=1.645(1)$
for $S=1/2$ and $9.194(3)$ for $S=3/2$ respectively.\cite{ganesh}

\section{In-plane frustration and the bismuth manganese oxynitrate materials}

Since the spin-$3/2$ material Bi$_3$Mn$_4$O$_{12}$NO$_3$ is unlikely
to have $J_\perp$ values an order of magnitude larger than in plane
couplings, it is clear that frustration must be present to 
explain the absence of N\'eel order. 
To study this we add a frustrating second neighbor
interaction $J_2$ to our study of the $S=3/2$ model. We find that the
addition of frustration rapidly decreases the critical value
of $J_\perp/J_1$. The estimated phase
boundary is shown in Fig.~6.
For $J_2/J_1=0.1$ this value is reduced to
$5.5\pm 1$, which is still significantly larger than the estimated ratio
between $1$ and $2$ for the material in the electronic structure calculation.
We find that a value of $J_2/J_1$ larger than $0.15$
can  lead to a disorder with $J_\perp/J_1$
less than $2$. These numbers are within the range of values estimated for
the material from electronic strusture calculations.\cite{kandpal} 

\begin{figure}
\begin{center}
 \includegraphics[width=6cm,angle=270]{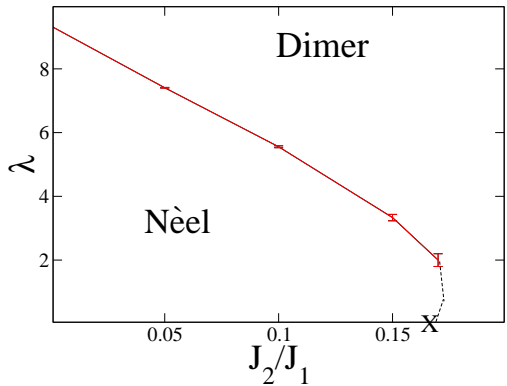}
\caption{\label{fig:Fig6} 
Phase Diagram for the $S=3/2$ model with frustration in the honeycomb layers.
}
\end{center}
\end{figure}

At the classical level, a single layer Heisenberg model
loses N\'eel order at $J_2/J_1=1/6$.\cite{fouet}
This is indicated by $X$ in Fig.~6.
However, this ratio is somewhat enhanced
in the presence of quantum fluctuations. Since, this enhancement
in the critical $J_2/J_1$ is roughly $0.03$ for the $S=1/2$ case,\cite{oitmaa-singh}
it should be much smaller for the $S=3/2$ case.
The critical $J_2/J_1$ should be further enhanced by the addition of 
a small $J_\perp$ in the
bilayer, since a small bilayer coupling makes the system more
ordered. Thus, the phase diagram in the $J_2/J_1$ and $J_\perp/J_1$ plane
has a slightly renetrant character for small $J_\perp/J_1$. 
This has the effect that beyond
$J_2/J_1=0.17$, the phase boundary becomes close to
vertical and our series analysis is no longer accurate to locate it precisely.
We sketch this phase boundary in Fig.~5 by dashed lines. We have analyzed the series
at $J_2/J_1=0.2$ and it shows no consistent point where the gap vanishes.

If the material Bi$_3$Mn$_4$O$_{12}$NO$_3$ has $J_2/J_1<0.17$, then the singlet
phase must arise from strong bilayer coupling and the material is almost
certainly in the spin-disordered phase that is adiabatically related to the
product singlet phase at large $J_\perp$. In this phase our calculation
of the excitation spectra should be accurate. Neutron scattering on
the material can help determine more precise exchange parameters.

However, it is possible that the material has $J_2/J_1>0.17$ and in this
case depending on how large $J_\perp/J_1$ is, a magnetically disordered phase may or
may not be adiabatically related to the product singlet phase found at large $J_\perp$. 
At $J_\perp/J_1<2$, there maybe a phase transition from one
singlet phase to another. 
Whether a new spin-liquid phases arises in the spin-$3/2$ model
at small $J_\perp/J_1$ with frustration, and how far down in $J_\perp/J_1$
the product singlet phase 
continues remains an open question, which deserves further study.
If there is a frustration dominated spin-liquid phase for $S=3/2$
at small $J_\perp$ values,
and that is the appropriate phase for the material,
that would make the material much more interesting. In this case,
neutron scattering may show an absence of triplon-like excitations.
Further experimental study is needed before more conclusions can be drawn.

\section{Conclusions}

In conclusion, in this paper we have used series expansion methods 
to study the ground state properties of $S=1/2$ and $S=3/2$
bilayer honeycomb lattice Heisenberg models. We find that 
an explanation for the material Bi$_3$Mn$_4$O$_{12}$NO$_3$
requires significant frustration in the honeycomb planes.
We also present results
for the evolution of the spectra with $\lambda=J_\perp/J_1$ in the
spin-disordered phase that is adiabatically related to the product singlet
phase at large $J_\perp$. 
Further study of the experimental system, especially its triplet
excitation spectra is necessary before further conclusions can
be drawn as to whether the material is in the product singlet phase 
or in a novel spin-liquid phase.

\begin{acknowledgements}
We would like to thank Arun Paramekanti for many valuable discussions.
This work is supported in part by NSF grant number  DMR-1004231.
\end{acknowledgements}


\end{document}